\documentclass[12pt,preprint]{aastex}
\usepackage{graphicx}

\shorttitle{Weak Lensing By Nearby Clusters}
\shortauthors{Kubo et al.}

\begin{document}


\title{The Sloan Nearby Cluster Weak Lensing Survey}


\author{Jeffrey M. Kubo\altaffilmark{1}, James Annis\altaffilmark{1}, Frances Mei Hardin\altaffilmark{2}, Donna Kubik\altaffilmark{1}, Kelsey Lawhorn\altaffilmark{2}, Huan Lin\altaffilmark{1}, Liana Nicklaus\altaffilmark{2}, Dylan Nelson\altaffilmark{4}, Ribamar R.R. Reis\altaffilmark{1}, Hee-Jong Seo\altaffilmark{1}, Marcelle Soares-Santos\altaffilmark{1,3}, Albert Stebbins\altaffilmark{1},Tony Yunker\altaffilmark{2}}
\altaffiltext{1}{Center for Particle Astrophysics, Fermi National Accelerator Laboratory, Batavia, IL 60510}
\altaffiltext{2}{Illinois Mathematics and Science Academy, 1500 W Sullivan Rd, Aurora, IL, 60506}
\altaffiltext{3}{Instituto de Astronomia, Geofisica e Ciencias Atmosfericas, Universidade de Sao Paulo}
\altaffiltext{4}{Department of Astronomy, University of California, 601 Campbell Hall, Berkeley, CA, 94720}


\begin{abstract}
We describe and present initial results of a weak lensing survey of nearby ($\rm{z}\lesssim0.1$) galaxy clusters in the Sloan Digital Sky Survey (SDSS).  In this first study, galaxy clusters are selected from the SDSS spectroscopic galaxy cluster catalogs of \citet{miller05} and \citet{berlind06}.  We report a total of seven individual low redshift cluster weak lensing measurements which include: A2048, A1767, A2244, A1066, A2199, and two clusters specifically identified with the C4 algorithm.  Our program of weak lensing of nearby galaxy clusters in the SDSS will eventually reach $\sim 200$ clusters, making it the largest weak lensing survey of individual galaxy clusters to date.
\end{abstract}





\section{Introduction}
The number density of galaxy clusters as function of mass and redshift is a well known probe of cosmological parameters \citep{haiman01}.  Several methods now exist to calibrate the cluster mass (optical dynamics, galaxy infall, X-ray, Sunyaev Zel'dovich, and weak or strong gravitational lensing) each of which have their own advantages and drawbacks.  Of particular interest to future optical imaging surveys is weak lensing which uses the shape distortion of background galaxies induced by the gravity of a foreground cluster to measure the cluster mass.  This lensing distortion is especially useful since it is independent of the dynamical state of the cluster and allows the cluster halo to be probed out to very large radii.  Much progress has been made in cluster lensing studies since its initial detection \citep{tyson90}, however surprisingly a recent compilation estimates that only $\sim150$ individual clusters have been studied with weak lensing \citep{dahle07}.  Typically these clusters are imaged deeply to obtain a sufficient lensing signal but usually only cover a limited area around each cluster, probing the cluster halo in the inner few Mpc's.  At low redshift in particular weak lensing has not been well studied since this requires imaging an area of $\sim 1^{\circ}$ or more around each cluster \citep{joffre00}.  Studying a large sample of clusters with weak lensing at low redshift allows for a direct comparison to other methods of mass calibration which have been well studied in the low redshift regime.

To date the largest area imaging survey is the Sloan Digital Sky Survey (SDSS) \citep{york00}.  Previous studies in the SDSS have measured weak lensing by stacking clusters \citep{sheldon08} in the redshift range $0.1<z<0.3$.  In spite of the shallow imaging, weak lensing of \emph{individual} clusters in the SDSS is also possible \citep{stebbins96,gould94} provided the clusters lie at low redshift.  In the low redshift limit $(z_{l}<<z_{s})$ the lensing shear signal is 
\begin{equation}\gamma_{t}\propto \frac{D_{l}D_{ls}}{D_{s}}\propto D_{l}\end{equation} where $D_{l}$ and $D_{s}$ are the angular diameter distances of the lens and source, and $D_{ls}$ is the angular diameter distance between the lens and source.  The lensing noise is 
\begin{equation}\delta\gamma_{t}\sim \sqrt{\rm{area}}\sim D_{l},\end{equation} so the lensing signal-to-noise ratio is equal to a constant.  Therefore by taking advantage of the large imaging area provided by the SDSS and ``going wide'' \citep{stebbins96} weak lensing can be used to probe the mass of low redshift galaxy clusters.  

Using the SDSS we recently reported a weak lensing measurement of the Coma Cluster \citep{kubo07} which is the lowest redshift cluster ($z=0.0236$) ever measured with weak lensing.  Since this study we have begun a program to measure other nearby, low redshift clusters in the SDSS.  In this letter we present the first results from our campaign using clusters selected from two publicly available spectroscopic galaxy cluster catalogs in the SDSS: the C4 catalog of \citet{miller05} and the \citet{berlind06} catalog.






\section{Data}
\label{sec:data}
\subsection{Imaging and Spectroscopy}
For our study we use data from the SDSS, an 8000 $\rm{deg}^{2}$ imaging and spectroscopic survey using a dedicated 2.5m telescope \citep{gunn06} at Apache Point Observatory in New Mexico.  Imaging is obtained in a time-delay-and-integrate (or drift scan) mode in five filters $ugriz$ \citep{fukugita96} using the SDSS imaging camera \citep{gunn98}.   The astrometric calibration of the SDSS is described in detail in \citet{pier03} and the photometric calibration pipeline is described in \citet{tucker06} and \citet{hogg01}. Targets for the SDSS spectroscopic survey are selected using automated algorithms described in \citet{strauss02} and spectra are obtained using two fiber-fed double spectrographs.  The main spectroscopic galaxy sample is complete to a magnitude of $r<17.77$.

\subsection{Source Galaxies}

In our lensing analysis we use galaxies drawn from the SDSS Data Release Six \citep{adelman08}.  Shape measurement is performed using the PHOTO pipeline \citep{lupton01} which measures the shapes of objects using adaptive moments \citep{bernstein02}.  We correct for the effects of the point spread function (PSF) using the linear PSF correction algorithm described in \citet{hirata03}.  Source galaxies used in our lensing analysis are required to be detected in each of the $ugriz$ bands, classified by PHOTO as galaxies (type=3), and have extinction corrected model magnitudes \citep{stoughton02} in the range $18<r<21.5$.  Other lensing studies in the SDSS have used more sophisticated star-galaxy classifiers \citep{sheldon08}, however similar to \citet{mandelbaum05} we use the PHOTO classification and restrict the sample to larger galaxies with a resolution factor $\rm{R}>0.33$ (or $1.5\times\rm{PSF}$).  Here R is given by
\begin{equation}\rm{R}=1-\frac{M^{\mathrm{PSF}}_{\mathrm{rrcc}}}{M_{\mathrm{rrcc}}}\end{equation}
where $M_{\rm{rrcc}}$ and $M_{\rm{rrcc}}^{\rm{PSF}}$ are the sum of the second order moments (in the CCD row and column directions) of the object and PSF respectively.  Our catalog is further restricted in that we only use shape measurements from the $r$ band which is the filter that typically has the best seeing \citep{adelman08}.  Photometric redshifts are drawn from the SDSS photoz2 table \citep{oyaizu08} and we restrict our sample to galaxies between $0.2<z_{\rm{phot}}<0.8$ with error $z_{\rm{err}}<0.4$.

\subsection{Cluster Sample}
\label{sec:clusters}

We have selected low redshift clusters in the SDSS from two publicly available spectroscopic catalogs: (1) the C4 cluster catalog of \citet{miller05} and (2) the \citet{berlind06} catalog.  The public C4 catalog is based on the SDSS Data Release Two \citep{abazajian04} and contains 748 clusters in the redshift range $0.02<z<0.17$ over a 2600 $\rm{deg}^{2}$ region.  The C4 algorithm searches for clusters in a seven dimensional parameter space which includes position ($\rm{R.A.}$,decl.), redshift, and four colors ($u-g$,$g-r$,$r-i$,$i-z$).  The use of galaxy colors in this algorithm is found to minimize cluster projection effects \citep{miller05}.  The public C4 catalog contains three different cluster centers which include the peak in the C4 density field, a luminosity-weighted mean centroid, and the position of the BCG (brightest cluster galaxy). 

The Berlind catalog uses the friends-of-friends algorithm of \citet{huchra82} to search for groups and clusters in the SDSS NYU Value Added Galaxy Catalog (NYU-VAGC) \citep{blanton05} which is equivalent to the SDSS Data Release Three \citep{abazajian05}.  The Berlind sample consists of three volume limited catalogs and we use the Mr20 (absolute r magnitude brighter than $-20$) catalog to search for clusters since it extends over the broadest redshift range $0.015<z<0.1$.  For each system the Berlind catalog measures several cluster parameters including an unweighted group centroid, a mean redshift, and a richness estimate.  For this study we select five of the richest clusters from the C4 catalog and two rich clusters from the Berlind catalog, described further in $\S \ref{sec:results}$.
 
\section{Weak Lensing Analysis}
\label{sec:lensing}
\subsection{Mass Model}

The lensing shear due to a cluster is given by 
\begin{equation}\gamma_{t}=\frac{\bar{\Sigma}(\leq r)-\Sigma(r)}{\Sigma_{\mathrm{crit}}}\end{equation}
where $\bar{\Sigma}(\leq r)$ is the average projected mass density within a radius r, $\Sigma(r)$ is the projected mass density at r, and $\Sigma_{\rm{crit}}$ is the critical surface mass density \citep{miralda91}.  We fitted the data for each cluster to a Navarro, Frenk, \& White profile (NFW) \citep{navarro96} which has been found in numerical simulations to provide an excellent description of dark matter halos, ranging from galaxy to cluster sized haloes.  From this mass profile a virial mass $(M_{200})$ can be determined from \begin{equation}M_{200}=\frac{800\pi}{3}\rho_{c}r_{200}^{3},\end{equation} where the virial radius $(r_{200})$ is defined as the radius where the density reaches a value of 200 times $\rho_{c}$ (the critical density of the universe at the redshift of the cluster).  The virial radius is related to the halo concentration ($c_{200}$) by $r_{200}=c_{200}/r_{s}$ where $r_{s}$ is the scale radius, the radius where the density profile changes shape \citep{navarro96}.  The expression for the tangential shear due to an NFW halo is given in \citet{wright00}.

To measure the shear in each cluster we use an unweighted shear estimator given by
\begin{equation}\gamma_{t}=\frac{1}{2\mathcal{R}}\frac{\sum e_{t}}{N}\label{eqn:shear},\end{equation}
where $e_{t}$ is the tangential ellipticty and N is the number of galaxies measured in logarithmically spaced annuli relative to the center of each cluster.  Our use of an unweighted estimator is valid since our source galaxies are in the magnitude and size range where the shape measurement error in each galaxy is small compared to the intrinsic shape noise.  We scaled the tangential ellipticity to a tangential shear in each bin using the shear responsivity $\mathcal{R}=1-\sigma_{\rm{SN}}^{2}$, where we assume a fixed value of $\sigma_{\rm{SN}}=0.37$ in all of our measurements \citep{hirata04}.

To fit the data to an NFW model we adopt a likelihood approach.   We define a binned likelihood given by
\begin{equation}\mathcal{L}=\prod_{i}^{N_{\rm{bins}}}\frac{1}{\sqrt{2\pi}}e^{\frac{-(\chi-\chi_{\rm{NFW}})^{2}}{2}}\end{equation}
where $\chi$ and $\chi_{\rm{NFW}}$ are 
\begin{equation}\chi=\frac{\bar{e}_{t}}{\sqrt{\frac{\sigma_{\bar{e}_{t}}}{N}}}, \quad \chi_{\rm{NFW}}=\frac{\bar{e}_{t}^{\rm{NFW}}}{\sqrt{\frac{\sigma_{\bar{e}_{t}}}{N}}}.\end{equation}
Here $\bar{e}_{t}$ is the mean tangential ellipticity and $\sigma_{\bar{e}_{t}}$ is the standard deviation of the mean tangential ellipticity.

\subsection{Results}
\label{sec:results}
\begin{figure}
\begin{tabular}{cc}
\includegraphics[height= 4 cm,width=7cm]{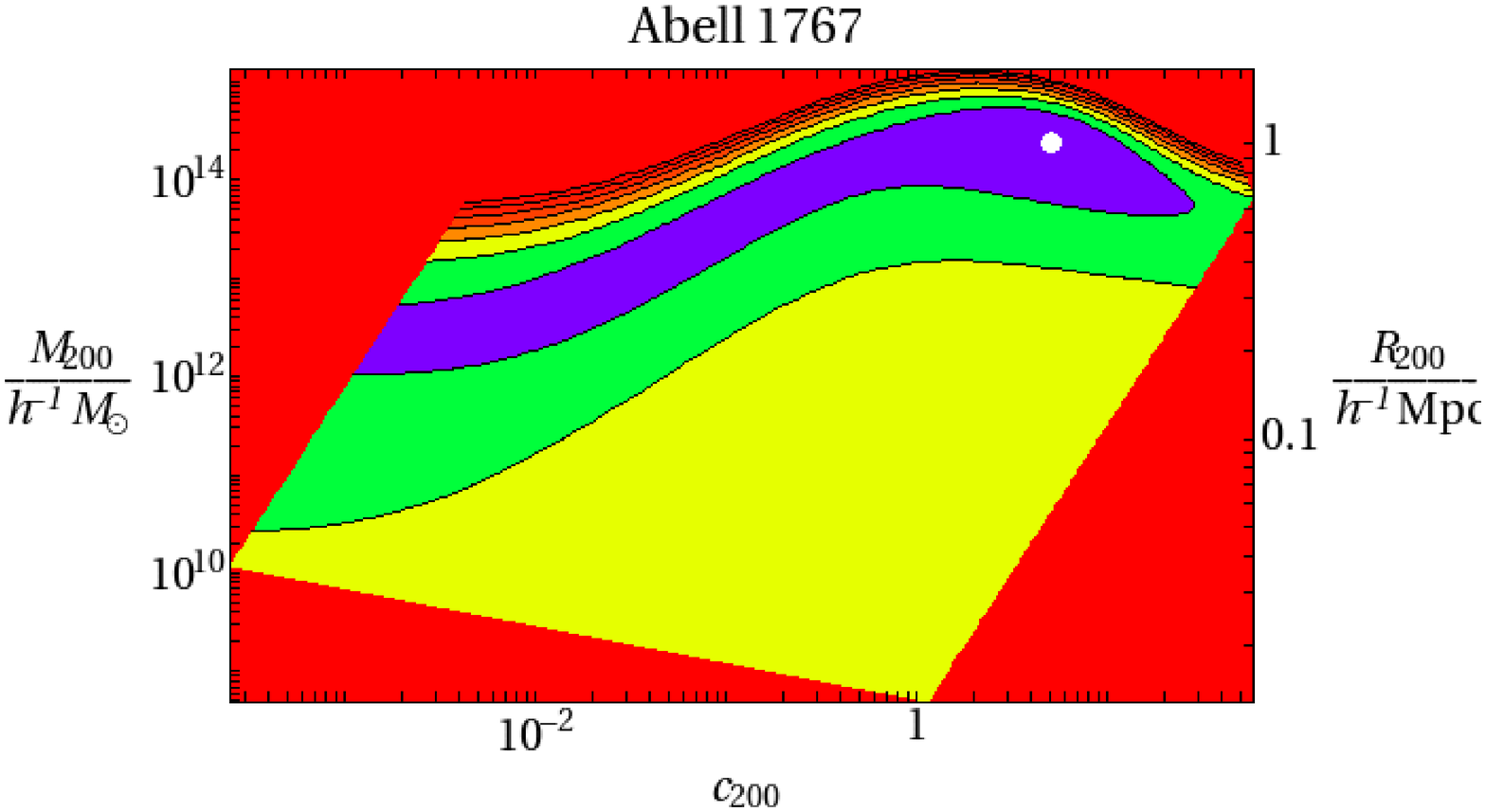} & 
\includegraphics[height= 4 cm,width=7cm]{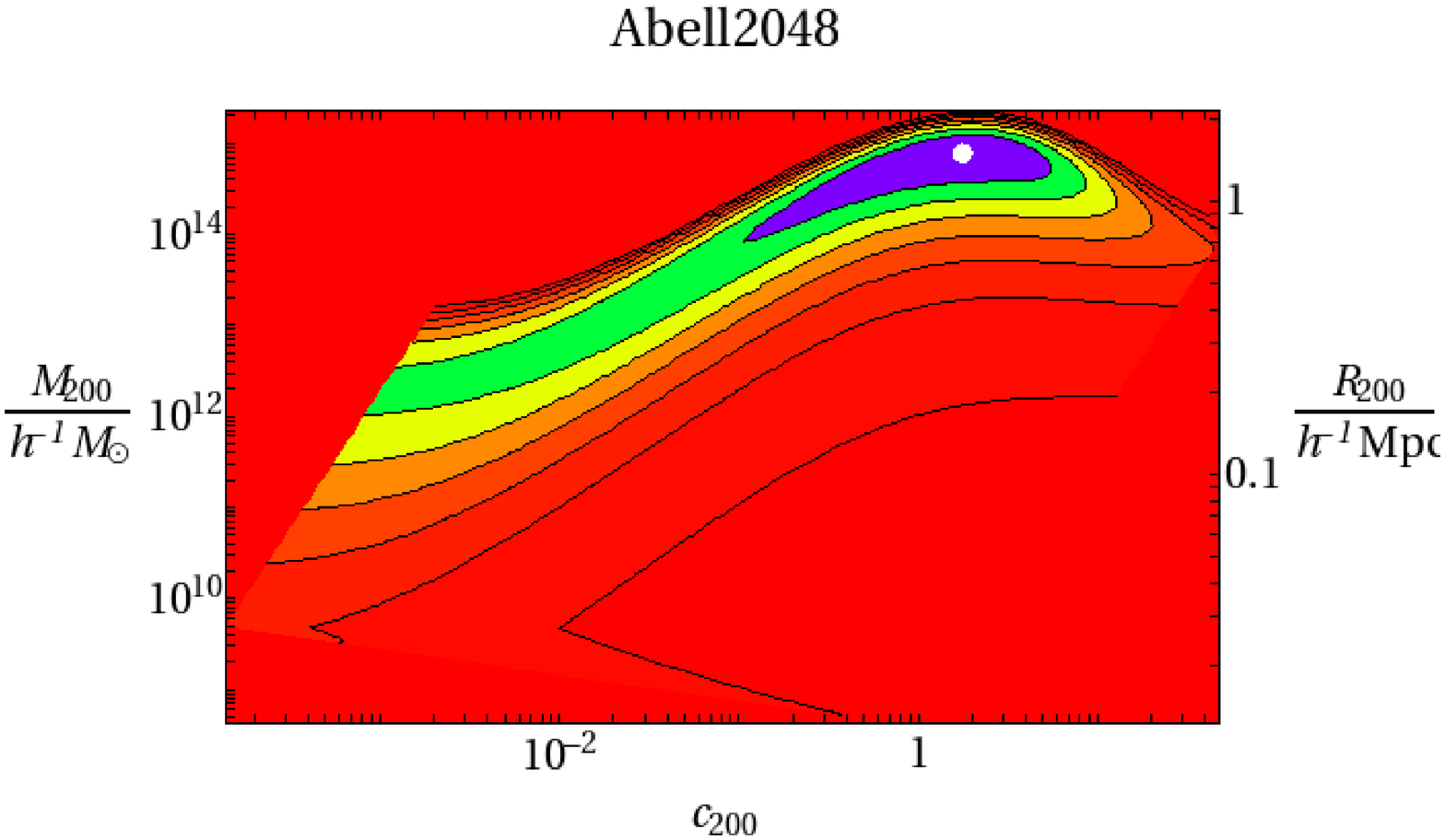} \\ 
\includegraphics[height= 4 cm,width=7cm]{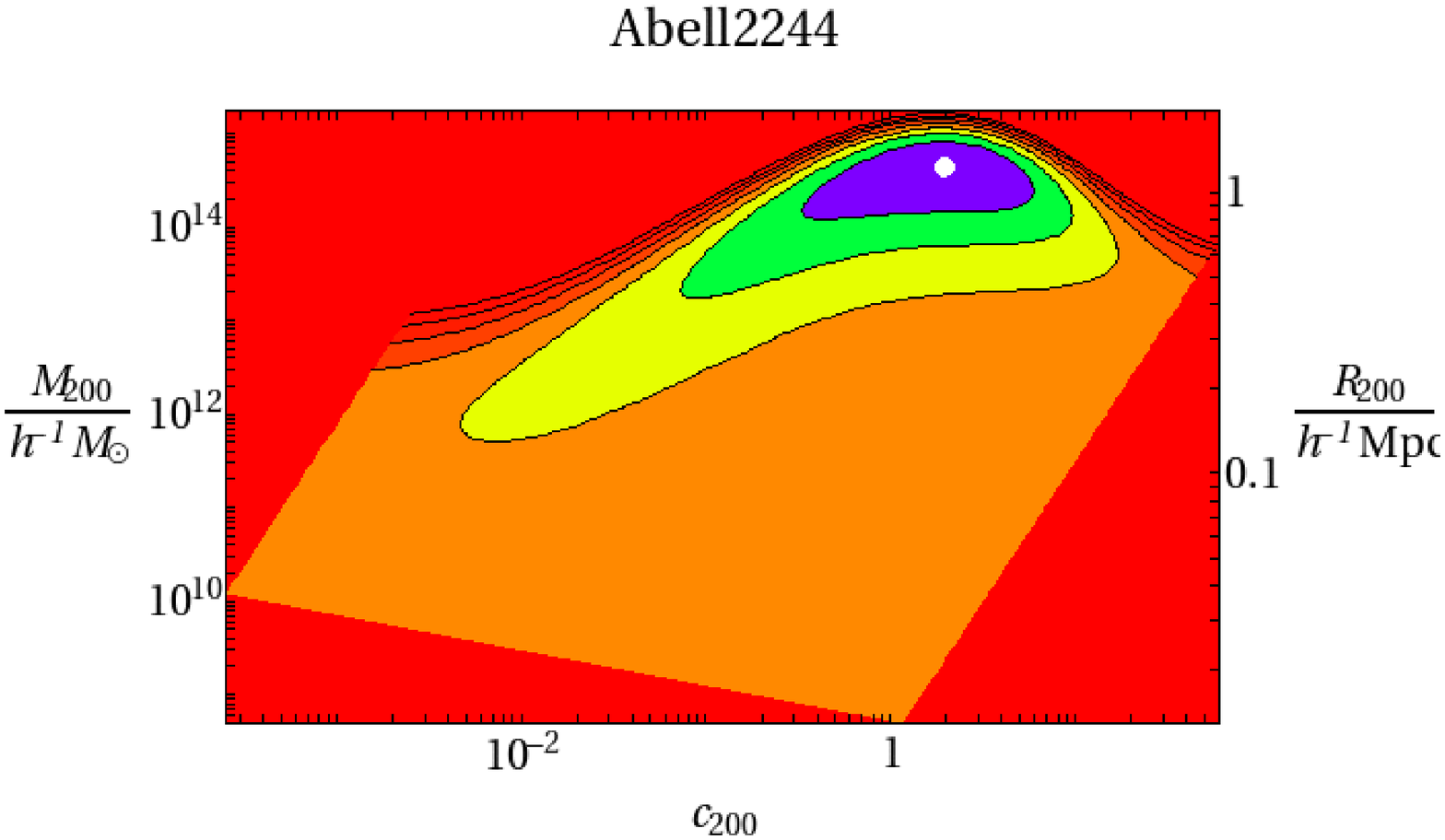} & 
\includegraphics[height= 4 cm,width=7cm]{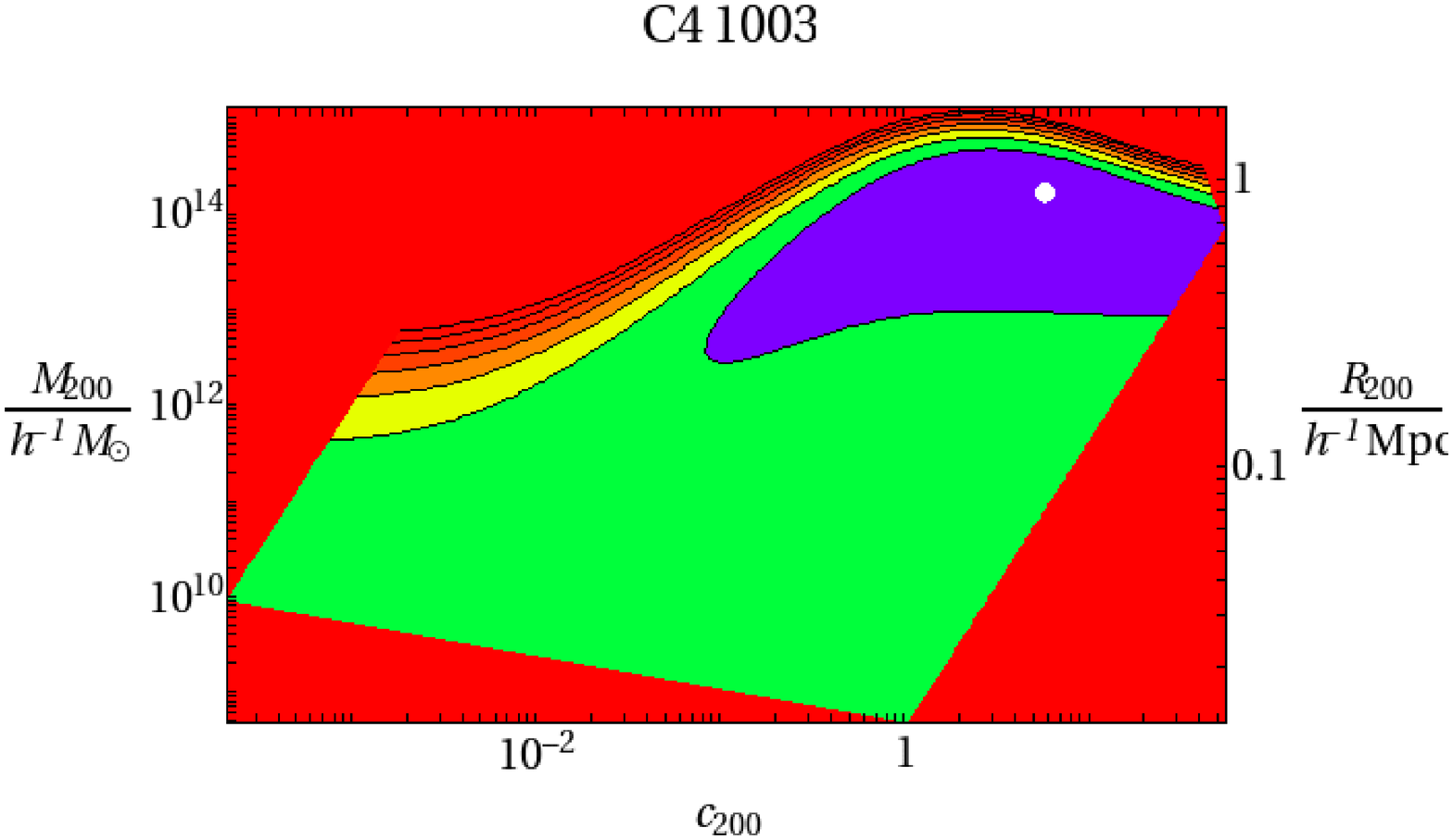} \\ 
\includegraphics[height= 4 cm,width=7cm]{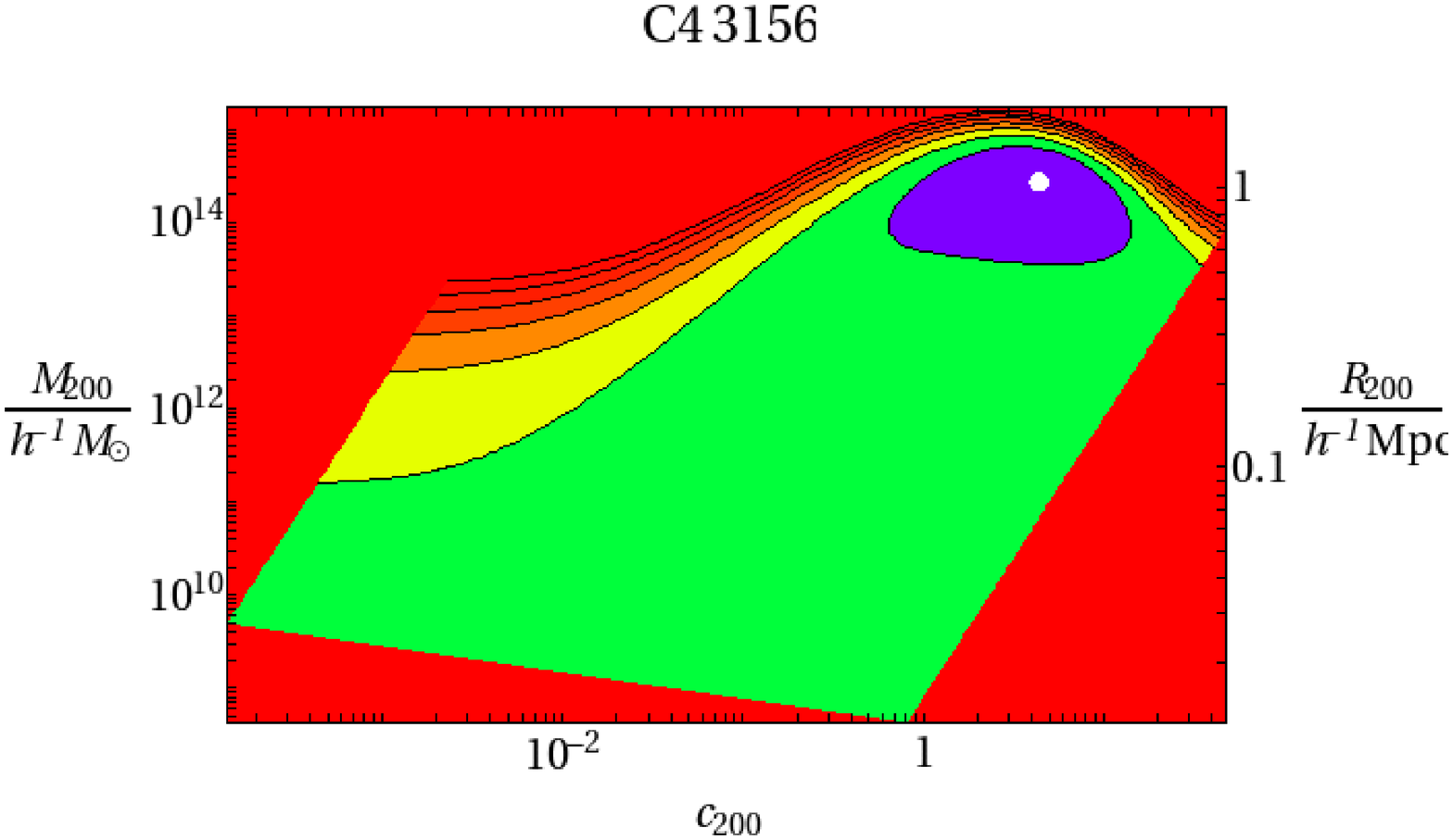} & 
\includegraphics[height= 4 cm,width=7cm]{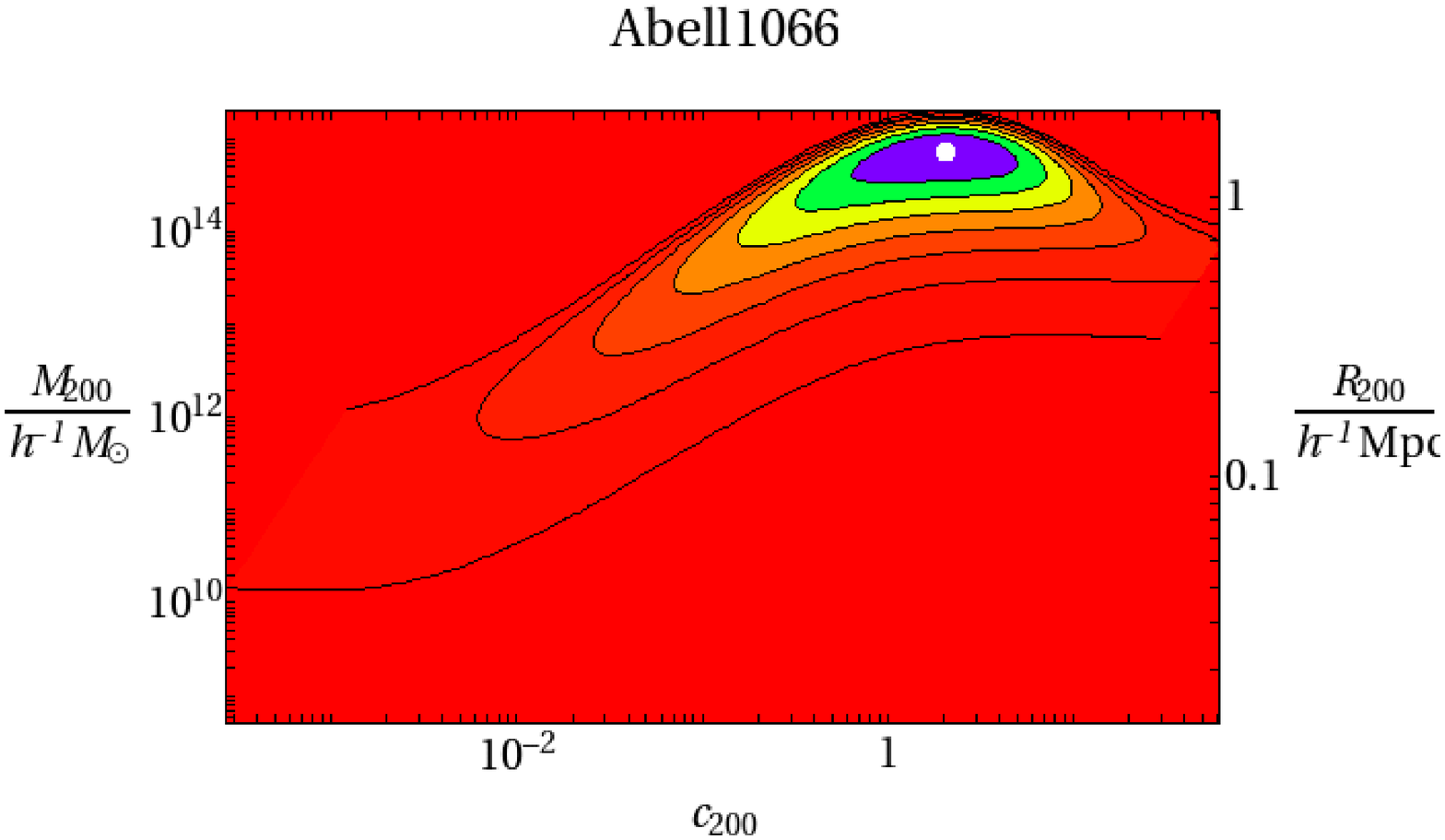} \\ 
\multicolumn{2}{c}{\includegraphics[height= 4 cm,width=7cm]{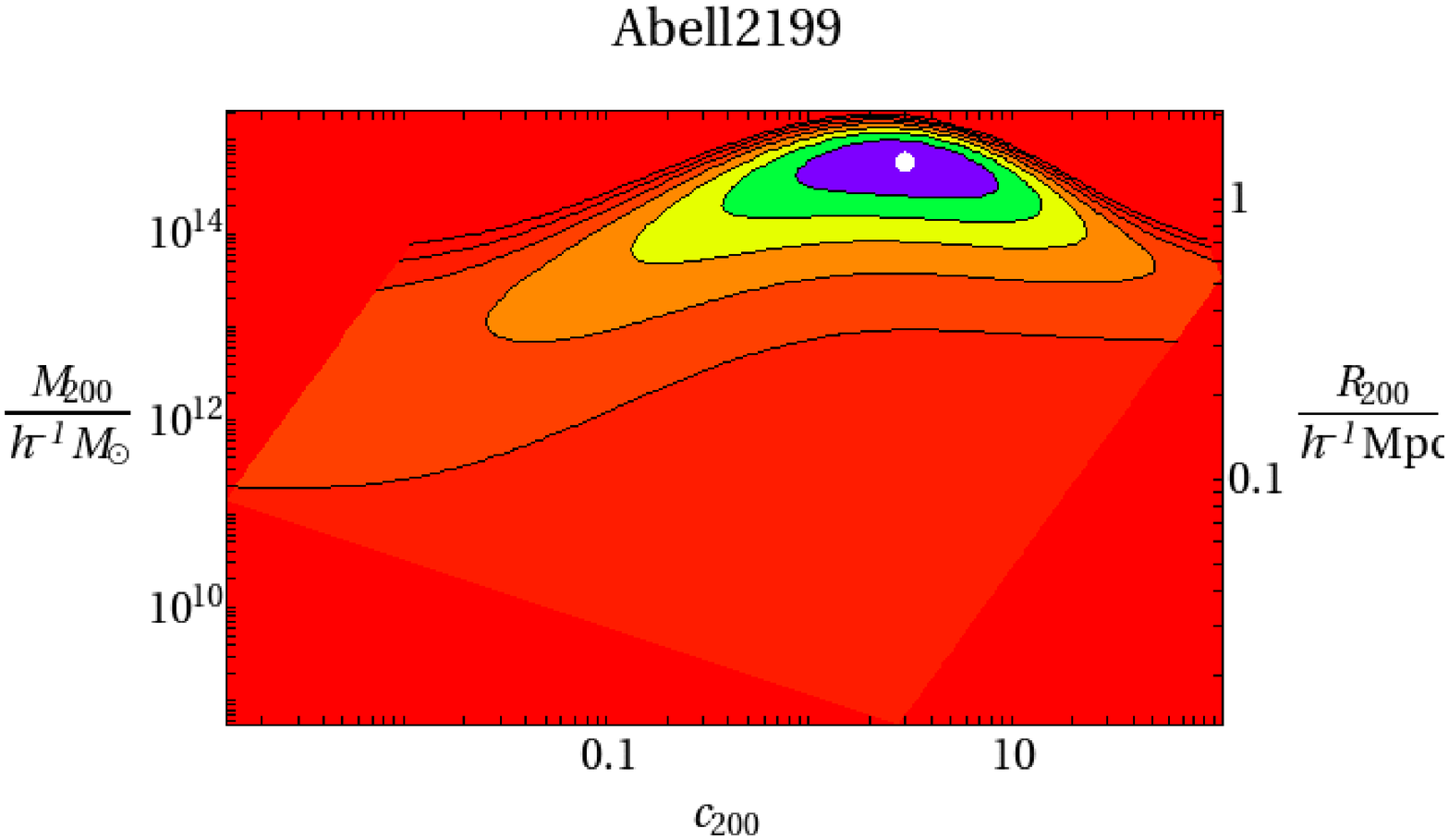}} \\
\end{tabular} 
\caption{Weak Lensing virial mass $(M_{200})$ and halo concentration ($c_{200}$) contours for each cluster in our sample. On the right hand side of each panel we also indicate the corresponding virial radius $r_{200}$.  The solid white dot in each panel is the maximum likelihood value and contours represent the $50\%$ (blue), $75\%$ (green), and $87.5\%$ (yellow) confidence regions.}
\label{fig:contours}
\end{figure}

In Figure \ref{fig:contours} we show lilkelihood plots in $M_{200}$ vs. $c_{200}$ for five C4 clusters (A2048, A1767, A2244, C4 1003, and C4 3156) and two clusters from the Berlind catalog (A1066 and A2199).  The corresponding value of $r_{200}$ is also shown on the right panels in the figure.  The maximum likelihood value for each cluster is indicated as the large white dot and contours represent the $50\%$ (blue), $75\%$ (green), and $87.5\%$ (yellow) confidence regions. For all of the clusters the halo concentration is not well constrained, which is in general typical of weak lensing measurements of clusters.  For both sets of clusters we find that the shear is maximized when centered on the cluster BCG.  For the C4 clusters we use the reported BCG center in the catalog but for the Berlind clusters we re-center on the cluster BCG by hand.  All cluster centers reported in Table \ref{tab:cluster} indicate the position of the cluster BCG used in the lensing analysis. 

Mass model parameters for our sample of clusters are summarized in Table \ref{tab:cluster}.  The weak lensing masses of our cluster sample varies from $\rm{log(M_{200})\sim14.2-14.8}$ with the cluster redshifts ranging from $z=0.0306-0.0990$.  Because of the very wide area imaging available in the SDSS the shear field due to each cluster can be measured out to large separation.  With this sample we are finding that the error in the mass is slightly reduced when extending the outer radius out to $\sim15\rm{h^{-1}}\rm{Mpc}$, and this is the outer cutoff radius used in our analysis.  Beyond this radius the mass determination does not appear to improve, and may be subject to the effects of large scale structure (see $\S \ref{sec:projection})$.  Stacked measurement of galaxy clusters have measured the mean shear field out to $\sim 30 \rm{Mpc}$ \citep{sheldon08}, but this is the first time we know of that the shear field of individual clusters have been measured out to this large of a separation.  We briefly comment on two clusters in our sample below:

\textbf{Abell 2199} is a known supercluster which also contains the cluster Abell 2197.  \citet{rines02} measured the mass of this cluster using the galaxy infall method and obtained a central mass of $3.2\times10^{14}\rm{h^{-1}}\rm{M_{\odot}}$ which is within $1\sigma$ of our measurement.  Our imaging data is not deep enough to separate the contributions of A2199 and A2197 therefore our mass should be interpreted as an estimate of mass of the combined system.

\textbf{Abell 2244} has a relatively small separation on the sky $(\sim30\arcmin)$ and in redshift $\Delta z\sim0.01$ to the cluster Abell 2245.  Because of the relative proximity of these two clusters we also cannot separate out the individual contributions and report only an estimate of the combined mass for the system.  In our calculation of the weak lensing mass we have assumed the redshift of Abell 2244 ($z=0.0990$).

\subsection{Projection effects}
\label{sec:projection}
Because weak lensing relies on the shape distortion induced on background galaxies, intervening clusters along the line of sight could cause additional error in the mass estimates of clusters \citep{hoekstra01}.  In the case of the SDSS the shape distortion due to these background clusters is expected to be small since the SDSS imaging is relatively shallow. 
This effect could however become non-negligible for future imaging surveys such as the Dark Energy Survey \citep{annis05}  and the Joint Dark Energy Mission \citep{aldering05}.  Foreground structure correlated with the cluster itself could also cause additional error as well as cause the mass estimates to be biased upward \citep{metzler99} \citep{cen97}.  Our sample of low-redshift clusters could in principle be used to study this effect, however a method to correct for this effect has not been developed so we leave this to future work.






\section{Summary}
\label{sec:summary} 
Here we summarize the main points of this work:

1. In addition to the Coma Cluster \citep{kubo07}, weak lensing measurements of other \emph{individual} nearby ($z<0.1$) clusters in the SDSS are possible. 

2. We present the first results from our survey which includes seven low redshift clusters selected from existing SDSS spectroscopic cluster catalogs.  All of these clusters have no previous weak lensing measurement.

3. With the very wide imaging area provided by the SDSS we are able to probe the shear field of individual clusters out to $\sim 15 \rm{h^{-1}}\rm{Mpc}$, further than other previous measurements.

The final spectroscopic cluster catalogs using these two cluster finding algorithms on the SDSS Data Release 7 \citep{abazajian08} should identify a large number of other low redshift clusters for which individual weak lensing measurements are possible.  We are also studying weak lensing on non-spectroscopic cluster catalogs including the Abell catalog \citep{abell89} and the Giriadi optical cluster catalog \citep{girardi98}.  Results of these analyses and a detailed discussion of lensing systematics will be presented in future papers.  When complete our survey should include weak lensing mass estimates for $>200$ clusters over the entire SDSS, making it the largest study of individual clusters with weak lensing.  This will allow for a detailed comparison of weak lensing derived cluster masses with masses using a variety of other probes at low redshift.

\acknowledgments

Funding for the SDSS and SDSS-II has been provided by the Alfred P. Sloan Foundation, the Participating Institutions, the National Science Foundation, the U.S. Department of Energy, the National Aeronautics and Space Administration, the Japanese Monbukagakusho, the Max Planck Society, and the Higher Education Funding Council for England. The SDSS Web Site is http://www.sdss.org/.  M. Soares-Santos and R. Reis are supported by the Brazilian National Research Council (CNPq).

\clearpage



 \begin{deluxetable}{cccccc}
\tablecolumns{6} 
\tablewidth{0pc}
\tablecaption{Cluster Data and Lensing Virial Masses} 
\tablehead{ 
\colhead{Name} & \colhead{ID\tablenotemark{a}}  & \colhead{R.A.} & \colhead{Decl.} & \colhead{z} & \colhead{$\rm{log(M_{200})}$}\\
\colhead{}  & \colhead{} &\colhead{(J2000)} & \colhead{(J2000)} & \colhead{} & \colhead{}\\
}
\startdata 
Abell 1767 & 3011 & 	204.0347 &  59.2064 &	0.0704 & $14.34^{+0.28}_{-0.54}$\\
Abell 2048 & 8129 &	228.8088 &  4.3862 &	0.0949 & $14.78^{+0.22}_{-0.32}$\\
Abell 2244 & 3004 &	255.6771 &  34.0600 &	0.0990 & $14.46^{+0.30}_{-0.56}$\\
C4 1003    & 1003 &   184.4213 &  3.6558 &    0.0771 & $14.20^{+0.36}_{-1.17}$\\
C4 3156      & 3156&    258.8017 &  64.3191 &	0.0950 & $14.34^{+0.34}_{-0.80}$\\
\tableline
Abell 1066 & 12289 &	159.7776 &  5.2098 &	0.0680 & $14.78^{+0.20}_{-0.30}$\\
Abell 2199 & 16089 &	247.1593 &  39.5512 &	0.0306 & $14.66^{+0.22}_{-0.32}$\\
\enddata 
\tablenotetext{a}{Cluster ID given in either the Berlind or C4 catalog.}
\tablecomments{The first five clusters in the table are from the C4 catalog, the last two are from the Berlind catalog. Errors bars on $\rm{log(M_{200})}$ are $1\sigma$ errors, where $M_{200}$ is in units of $\rm{h^{-1}M_{\odot}}$.}
\label{tab:cluster}
\end{deluxetable}

\clearpage



\clearpage



\clearpage


\end{document}